\documentclass [a4paper,fleqn, 12pt]{article}
\usepackage{graphicx}

\usepackage {amsmath} \usepackage{amssymb} \usepackage{cite}

\newcommand{\cn}

\begin{document}

\title{On connections of the Li\'{e}nard equation with some equations of  Painlev\'{e}--Gambier type}

\author{Nikolay A. Kudryashov, \and Dmitry I. Sinelshchikov}

\date{Department of Applied Mathematics, National Research Nuclear University MEPHI, 31 Kashirskoe Shosse, 115409 Moscow, Russian Federation}

\maketitle

\begin{abstract}
The Li\'{e}nard equation is used in various applications. Therefore, constructing general analytical solutions of this equation is an important problem.
Here we study connections between the Li\'{e}nard equation and some equations from the Painlev\'{e}--Gambier classification. We show that with the help of such connections one can construct general analytical solutions of the Li\'{e}nard equation's subfamilies. In particular, we find three new integrable families of the  Li\'{e}nard equation. We also propose and discuss an approach for finding one--parameter families of closed--form analytical solutions of the Li\'{e}nard equation.
\end{abstract}


\section{Introduction}

We consider the Li\'{e}nard equation
\begin{equation}
y_{zz}+f(y)y_{z}+g(y)=0,
\label{eq:L1}
\end{equation}
where $f$ and $g$ are arbitrary functions, which do not vanish simultaneously. Equation \eqref{eq:L1} has a vast range of applications in physics, biology and other fields of science \cite{Holmes1983,Andronov,Polyanin,Lakshmanan2003}.

The Li\'{e}nard equation has been thoroughly investigated from a dynamical systems point of view (see, e.g. \cite{Villari1987,Perko,Depassier2001,Villari2005} and references therein). However, only a few studies were devoted to the construction of closed--form analytical solutions of the Li\'{e}nard equation. For instance, Lie point symmetries of equation \eqref{eq:L1} were studied in \cite{Bluman2008,Lakshmanan2009,Lakshmanan2009a} and equations having eight, three and two parameters symmetries, i.e. those equations that can be integrated by the Lie method, were found. Integrability of the Li\'{e}nard equation with the help of the Prelle--Singer method was considered in \cite{Lakshmanan2005} and some integrable Li\'{e}nard equations were constructed. Authors of \cite{Meleshko2010,Meleshko2011} found necessary and sufficient conditions for linearization of the Li\'{e}nard--type equation via the generalized Sundman transformations. In \cite{Kudryashov2016} it was shown that integrability conditions for equation \eqref{eq:L1} obtained with the help of the Chiellini lemma (see, e.g. \cite{Mancas2013,Harko2014}) are equivalent to linearizabiliy conditions via the generalized Sundman transformations. Moreover, authors of \cite{Kudryashov2016} also pointed out that equation \eqref{eq:L1} with a maximal point symmetries group can be linearized by the generalized Sundman transformations.


Recently in \cite{Kudryashov2014,Kudryashov2015,Kudryashov2015a,Kudryashov2016} it has been shown that new families of integrable Li\'{e}nard equations can be constructed if one considers connections between the Li\'{e}nard--type equations and some other nonlinear differential equations, which have the closed--form general analytical solutions. Let us remark that both in \cite{Kudryashov2014,Kudryashov2015,Kudryashov2015a,Kudryashov2016} and in this work it is supposed that such connections are given by the generalized Sundman transformations. For example, authors of \cite{Kudryashov2015a} proved that the quadratic Li\'{e}nard equation (see, e.g. \cite{Sabatini2004,Nucci2015}) can be mapped into an equation for the elliptic functions via the generalized Sundman transformations for arbitrary coefficients. In \cite{Kudryashov2016} a new criterion for the integrability of equation \eqref{eq:L1} was proposed using a connection between the Li\'{e}nard equation and a sub--case of equation \eqref{eq:L1} which general solution can be expressed via the Jacobi elliptic cosine.  Here we extend results of \cite{Kudryashov2016} and study connections between equation \eqref{eq:L1} and its sub--cases that are of the Painlev\'{e}--Gambier type (see, e.g. \cite{Ince}). As a result, we find three new families of integrable Li\'{e}nard equations. We also propose an approach for constructing one--parameter families of analytical solutions of equation \eqref{eq:L1}. To the best of our knowledge our results are new.


The rest of this work is organized as follows. In the next section we study connections between the Li\'{e}nard equation and some equations of the Painlev\'{e}--Gambier type and obtain three new criteria for integrability of the Li\'{e}nard equation. We also discuss an approach for constructing one--parameter families of analytical solutions of equation \eqref{eq:L1}. We demonstrate effectiveness of our approaches by several new examples of integrable Li\'{e}nard equations in section 3. In the last section we briefly discuss our results.


\section{Main results}

We study connections of the Li\'{e}nard equation with some equations from the Painlev\'{e}--Gambier classification that are given by the nonlocal transformations
\begin{equation}
w=F(y), \quad d\zeta=G(y)dz, \quad F_{y}G\neq0,
\label{eq:L1_1}
\end{equation}
where $w$ and $\zeta$ are new dependent and independent variables correspondingly. Transformations \eqref{eq:L1_1} are called the generalized Sundaman transformations (see, e.g. \cite{Meleshko2010,Meleshko2011}).

Let us note that throughout this work we consider only autonomous equations from the Painlev\'{e}--Gambier classification. Thus, when we discuss a certain Painlev\'{e}--Gambier type equation with variable coefficients we assume that these coefficients are constants. There are seven subcases of equation \eqref{eq:L1} with $f(y)\not\equiv0$ that belong to the Painlev\'{e}--Gambier classification \cite{Ince}. They are non--canonical forms of equations II, V and VII and equations VI, X, XXIV and XVII from Ince's book \cite{Ince} (in the last two equations it is assumed that the parameter $m$ is equal to 1).

It can be seen that equations VI, X and XXIV can be linearized by transformations \eqref{eq:L1_1} (see \cite{Meleshko2010,Kudryashov2016}). Therefore, equations from \eqref{eq:L1} which can be transformed into VI, X and XXIV by means of \eqref{eq:L1_1} can be linearized via \eqref{eq:L1_1} since a combination of Sundman transformations is a Sundman transformation. A connection between \eqref{eq:L1} and equation VII has been recently studied in \cite{Kudryashov2016}. Consequently, it is necessary to study connections between \eqref{eq:L1} and equations II, V and XVII. These connections lead us to new criteria for integrability of the Li\'{e}nard equations.


\subsection{Equation II from the Painlev\'{e}--Gambier classification}
First of all, we consider the following non--canonical form of equation II from the Painlev\'{e}--Gambier classification \cite{Ince}
\begin{equation}
w_{\zeta\zeta}+5w_{\zeta}+6w-w^{2}=0.
\label{eq:L1_3}
\end{equation}
The general solution of equation \eqref{eq:L1_3} can be written as follows
\begin{equation}
w=6e^{-2(\zeta-\zeta_{0})}\wp\left\{e^{-(\zeta-\zeta_{0})},0,g_{3}\right\},
\label{eq:L1_5}
\end{equation}
where $\wp$ is the Weierstrass elliptic function and $g_{3}$ is an arbitrary constant. Let us note that throughout this work we denote by $\zeta_{0}$ an arbitrary constant corresponding to the invariance of the studied equations under shift transformations in an independent variable.

There is a non--autonomous first integral of equation \eqref{eq:L1_3}:
\begin{equation}
6(w_{\zeta}+2w)^{2}-4w^{3}=216g_{3}e^{-6\zeta}.
\label{eq:L1_7}
\end{equation}
Below we show that relation \eqref{eq:L1_7} can be used for the construction of one--parametric analytical solutions of the Li\'{e}nard equation. Notice also that solution \eqref{eq:L1_5} degenerates in the case of $g_{3}=0$ and can be found from \eqref{eq:L1_5}:
\begin{equation}
w=6(e^{\zeta-\zeta_{0}}+1)^{-2}.
\label{eq:L1_5_a}
\end{equation}
Now we discuss a connection between the whole family of the Li\'{e}nard equations and equation \eqref{eq:L1_5}.

\textbf{Theorem 1}. Equation \eqref{eq:L1} can be transformed into \eqref{eq:L1_3} by means of \eqref{eq:L1_1} with
\begin{equation}
F=\kappa\left(\int f dy+\lambda\right), \quad G(y)=\frac{1}{5}f,
\label{eq:L1_9}
\end{equation}
if the following correlation on functions $f$ and $g$ holds
\begin{equation}
g=-\frac{f}{25}\left[\int f dy +\lambda\right]\left[\kappa \left(\int f dy+\lambda\right)-6\right],
\label{eq:L1_11}
\end{equation}
where $\kappa\neq0$ and $\lambda$ are arbitrary parameters. \\
\textbf{Proof.} Using transformations \eqref{eq:L1_1} we can express derivatives of $y$ with respect to $z$ via derivatives of $w$ with respect to $\zeta$. Then we substitute these expressions into equation \eqref{eq:L1} and require that the result is equation \eqref{eq:L1_3}. As a consequence, we obtain a system of two ordinary differential equations on functions $F$ and $G$ and a correlation on functions $f(y)$ and $g(y)$. Solving these equations with respect to $F$, $G$ and $g$ we get formulas \eqref{eq:L1_9} and \eqref{eq:L1_11}. This completes the proof.

Since equation \eqref{eq:L1} under condition \eqref{eq:L1_11} is connected with \eqref{eq:L1_3} via \eqref{eq:L1_1} one can suppose that we can obtain a first integral of \eqref{eq:L1} under \eqref{eq:L1_11} form first integral \eqref{eq:L1_7} of \eqref{eq:L1_3}. However, first integral \eqref{eq:L1_7} is non--autonomous and we can obtain a first order integro--differential equation for solutions of equation \eqref{eq:L1} under condition \eqref{eq:L1_11}. Therefore, we get the following consequence of Theorem 1.

\textbf{Corollary 1}. If condition \eqref{eq:L1_11} holds, then solutions of equation \eqref{eq:L1} satisfy the following relation
\begin{equation}
\begin{gathered}
6\kappa^{2}\left[5y_{z}+2\left(\int f dy+\lambda\right)\right]^{2}-4\kappa^{3}\left(\int f dy+\lambda\right)^{3}=216e^{-6\zeta}g_{3},\\
\zeta=\frac{1}{5}\int\limits_{0}^{z}f[y(\xi)]d\xi,
\label{eq:L1_9_a}
\end{gathered}
\end{equation}
Note that here and below we denote by $\xi$ a dummy integration variable.

At first glace it may seen that relation \eqref{eq:L1_9_a} is not useful. However, assuming that $g_{3}=0$ in \eqref{eq:L1_9_a} we obtain a first order ordinary differential equation the general solution of which gives us a one--parameter solutions family  of equation \eqref{eq:L1}. In the next section we will demonstrate applications of Theorem 1 and Corollary 1 for finding closed--form analytical solutions of the Li\'{e}nard equation.


\subsection{Equation V from the Painlev\'{e}--Gambier classification}
The next equation that we consider is a non--canonical form of number V equation from the Painlev\'{e}--Gambier classification (see \cite{Ince}, p. 331)
\begin{equation}
w_{\zeta\zeta}+2ww_{\zeta}-\alpha(w_{\zeta}+w^{2})+\beta=0,
\label{eq:L1_V}
\end{equation}
where $\alpha\neq0$, $\beta$ are arbitrary parameters.

The general solution of equation \eqref{eq:L1_V} has the form
\begin{equation}
w=\frac{\Psi_{\zeta}}{\Psi}, \,\,\, \Psi=C_{2}I_{\chi}\left(\frac{2\sqrt{C_{1}}}{\alpha}e^{\alpha\zeta/2}\right)+C_{3}K_{\chi}\left(\frac{2\sqrt{C_{1}}}{\alpha}e^{\alpha\zeta/2}\right), \,\,\, \chi^{2}=\frac{4\beta}{\alpha^{3}},
\label{eq:L1_V_gs}
\end{equation}
where $I_{\chi}$, $K_{\chi}$ are the modified Bessel functions, $C_{1}$ and $C_{2}^{2}+C_{3}^{2}\neq0$ are arbitrary constants. Let us remark that in fact solution \eqref{eq:L1_V_gs} depends only on two arbitrary constants and without loss of generality either $C_{2}$ or $C_{3}$ can be set equal to 1.

Equation \eqref{eq:L1_V} admits the following non--autonomous first integral
\begin{equation}
w_{\zeta}+w^{2}-\frac{\beta}{\alpha}=C_{1}e^{\alpha\zeta}.
\label{eq:L1_V_fi}
\end{equation}
Thus, using \eqref{eq:L1_V_fi} we get a special solution of equation \eqref{eq:L1_V} corresponding to the case of $C_{1}=0$:
\begin{equation}
w=\sqrt{\frac{\beta}{\alpha}}\tanh\left\{\sqrt{\frac{\beta}{\alpha}}(\zeta-\zeta_{0})\right\}.
\label{eq:L1_V_ps}
\end{equation}
Notice that in the case of $\beta=0$ one--parameter family of solutions of \eqref{eq:L1_V} has the form $w=(\zeta-\zeta_{0})^{-1}$.

Now we are in position to obtain a criterion of equivalence between equations \eqref{eq:L1} and \eqref{eq:L1_V}.

\textbf{Theorem 2}. Suppose that the following correlation on functions $f$ and $g$ holds
\begin{equation}
\begin{gathered}
g=\mp\frac{f}{\nu\sqrt{\mu+\nu\int f dy }}\left(\alpha\left[\alpha\pm\sqrt{\mu+\nu\int f dy }\right]^{2}-4\beta\right),
\label{eq:L1_V_3}
\end{gathered}
\end{equation}
where $\mu$ and $\nu\neq0$ are arbitrary parameters; then equation \eqref{eq:L1} can be transformed into \eqref{eq:L1_V} by means of \eqref{eq:L1_1} with
\begin{equation}
\begin{gathered}
F=\frac{1}{2}\left(\alpha\pm\sqrt{\mu+\nu\int f dy }\right), \quad
G=\pm\frac{f}{\sqrt{\mu+\nu\int f dy }}.
\label{eq:L1_V_1}
\end{gathered}
\end{equation}
\textbf{Proof.} The proof is similar to that of Theorem 1, and is omitted.

As a straightforward consequence of Theorem 2 we obtain the following result.

\textbf{Corollary 2}. Suppose that correlation \eqref{eq:L1_V_3} holds; then solutions of equation \eqref{eq:L1} satisfy the following relation
\begin{equation}
\frac{\nu}{4}y_{z}+\left(\frac{\alpha}{2}\pm\frac{1}{2}\sqrt{\mu+\nu\int f dy }\right)^{2}-\frac{\beta}{\alpha}=C_{1}\exp\left\{\pm\alpha\int_{0}^{z}G(y(\xi))d\xi\right\},
\label{eq:L1_V_5}
\end{equation}
where $G$ is given by \eqref{eq:L1_V_1}.

One can find special solutions of equation \eqref{eq:L1} under condition \eqref{eq:L1_V_3} if we assume that $C_{1}=0$ in \eqref{eq:L1_V_5} and solve corresponding first order differential equation. Applications of Theorem 2 and Corollary 2 will be considered in the next section.

\subsection{Equation XXVII from the Painlev\'{e}--Gambier classification}

Now we consider a special case of equation $\mbox{XXVII}$ from the Painlev\'{e}--Gambier classification (see \cite{Ince}, p. 338 ). We suppose that $m=1$, $f=\phi=0$ and $\psi=\gamma$,  where $\gamma\neq0$ is an arbitrary parameter.  Then equation $\mbox{XXVII}$ takes the form
\begin{equation}
w_{\zeta\zeta}-\frac{1}{w} w_{\zeta}-\gamma w+\frac{1}{w}=0.
\label{eq:L1_XXVII}
\end{equation}
Note that connection between equation \eqref{eq:L1} and the general case of equation $\mbox{XXVII}$ at $m=1$ will be considered elsewhere.

The general solution of equation \eqref{eq:L1_XXVII} is given by
\begin{equation}
w=\frac{1}{\sqrt{\gamma}}\cosh\{\sqrt{\gamma}(\zeta-\zeta_{0})\}\bigg[\mbox{arctan}\big(\sinh\{\sqrt{\gamma}(\zeta-\zeta_{0})\}\big)+C_{4}\bigg],
\label{eq:L1_XXVII_gs}
\end{equation}
where $C_{4}$ is an arbitrary constant.

We can write a first integral of equation \eqref{eq:L1_XXVII} as follows
\begin{equation}
\frac{w_{\zeta}-1}{w}=\sqrt{\gamma}\frac{e^{\sqrt{\gamma}\zeta}-C_{5}e^{-\sqrt{\gamma}\zeta}}{e^{\sqrt{\gamma}\zeta}+C_{5}e^{-\sqrt{\gamma}\zeta}},
\label{eq:L1_XXVII_fI}
\end{equation}
where $C_{5}$ is an arbitrary constant, which is connected with $\zeta_{0}$.

In the case of $C_{5}=0$ corresponding special solution of equation \eqref{eq:L1_XXVII} has the form
\begin{equation}
w=e^{\sqrt{\gamma}(\zeta-\zeta_{0})}-\frac{1}{\sqrt{\gamma}}.
\end{equation}

Now we consider a connection between \eqref{eq:L1} and equation \eqref{eq:L1_XXVII}.

\textbf{Theorem 3}. Equation \eqref{eq:L1} can be transformed into \eqref{eq:L1_XXVII} by means of \eqref{eq:L1_1} with
\begin{equation}
\begin{gathered}
F=\sigma\exp\left\{\eta\int f dy\right\}, \quad
G=-\sigma f\exp\left\{\eta\int f dy\right\},
\label{eq:L1_XXVII_f}
\end{gathered}
\end{equation}
if the following correlation on functions $f$ and $g$ holds
\begin{equation}
\begin{gathered}
g=-\frac{f}{\eta}\left(\sigma^{2}\gamma \exp\left\{2\eta\int f dy\right\} -1\right),
\label{eq:L1_XXVII_c}
\end{gathered}
\end{equation}
where $\sigma\neq0$ and $\eta\neq0$ are arbitrary parameters.

\textbf{Proof.} The proof is similar to that of Theorem 1, and is omitted.

Theorem 3 leads us to the following statement.

\textbf{Corollary 3}. If correlation \eqref{eq:L1_XXVII_c} holds, then solutions of equation \eqref{eq:L1} satisfy the relation
\begin{equation}
\begin{gathered}
 \exp\left\{-\eta\int f dy\right\}\left(\eta y_{z}+1\right)=
-\sigma\sqrt{\gamma}\frac{e^{\sqrt{\gamma}\zeta}-C_{5}e^{-\sqrt{\gamma}\zeta}}{e^{\sqrt{\gamma}\zeta}+C_{5}e^{-\sqrt{\gamma}\zeta}},\\
\zeta=\int\limits_{0}^{z}G(y(\xi))d\xi.
\label{eq:L1_XXVII_a}
\end{gathered}
\end{equation}
Here $G$ is given by \eqref{eq:L1_XXVII_f}.

In the case of $C_{5}=0$ from \eqref{eq:L1_XXVII_a} we get an ordinary differential equation which general solution gives us a one--parametric family of solutions of equation \eqref{eq:L1} under correlation \eqref{eq:L1_XXVII_c}.

Finally, it is worth noting that criteria for the integrability of equation \eqref{eq:L1} obtained in Theorems 1--3 do not coincide with previously known integrability conditions for the Li\'{e}nard equation. Therefore, formulas \eqref{eq:L1_11}, \eqref{eq:L1_V_3} and \eqref{eq:L1_XXVII_c} give us new families of integrable Li\'{e}nard equations. Note also that with the help of results form \cite{Bluman2008,Lakshmanan2009,Lakshmanan2009a} one can see that equation \eqref{eq:L1} under conditions \eqref{eq:L1_11}, \eqref{eq:L1_V_3} and \eqref{eq:L1_XXVII_c} cannot be either integrated or linearized by the Lie method.

\section{Examples}

In this section we consider applications of above obtained connections between the Li\'{e}nard equation and equations of the Painlev\'{e}--Gambier type. We use these results for constructing both general and particular solutions of several members of equations family \eqref{eq:L1}.

\subsection{Example of Theorem 1 application.}

\begin{figure}[!h]
\center
\includegraphics[width=0.99\textwidth]{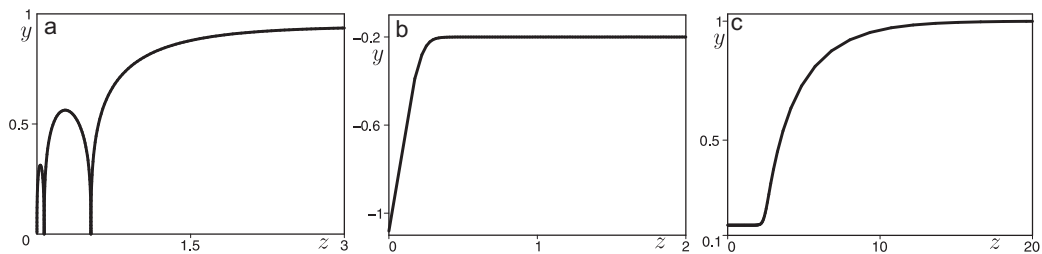}
\caption{Solution \eqref{eq:II_s_1} at: a) $a=\lambda=1$, $\kappa=-100$, $g_{3}=1$  and $\zeta_{0}=2.2$; b) $a=\lambda=\kappa=1$, $g_{3}=-1$  and $\zeta_{0}=0.5$; c) $a=\lambda=1$, $\kappa$=-1, $g_{3}=0$ and $\zeta_{0}=25$.  }
\label{f1}
\end{figure}

Let us consider the case of $f(y)=a/y^2$, where $a\neq0$ is an arbitrary parameter. With the help of Theorem 1 we find that
\begin{equation}
F(y)=-\frac{a\kappa}{y}+\kappa\lambda, \quad G(y)=\frac{a}{5y^{2}}.
\label{eq:LE_e1}
\end{equation}
Using \eqref{eq:L1} and \eqref{eq:L1_11} we find corresponding Li\'{e}nard equation
\begin{equation}
y_{zz}+\frac{a}{y^{2}}y_{z}-\frac{a}{25y^{4}}\left(\lambda y -a\right)\left([\lambda\kappa-6]y-a\kappa\right)=0.
\label{eq:LE_e1_1}
\end{equation}
Taking into consideration formulas \eqref{eq:L1_1}, \eqref{eq:L1_5}, \eqref{eq:LE_e1} we get the general solution of equation \eqref{eq:LE_e1_1}
\begin{equation}
\begin{gathered}
y=\frac{a\kappa}{\lambda\kappa-6e^{-2(\zeta-\zeta_{0})}\wp\left\{e^{-(\zeta-\zeta_{0})},0,g_{3}\right\}}, \quad
z=\frac{5}{a}\int y^{2} d\zeta.
\label{eq:II_s_1}
\end{gathered}
\end{equation}
In the degenerated case, i.e. when $g_{3}=0$, solution \eqref{eq:II_s_1} becomes
\begin{equation}
y=\frac{a\kappa(e^{\zeta-\zeta_{0}}+1)^{2}}{\kappa\lambda (e^{\zeta-\zeta_{0}}+1)^{2}-6}, \quad
z=\frac{5}{a}\int y^{2} d\zeta.
\label{eq:II_s_1_a}
\end{equation}
We demonstrate plots of solutions \eqref{eq:II_s_1}, \eqref{eq:II_s_1_a} in Fig.\ref{f1} (plates a,b and plate c correspondingly). We see that these solutions describe various kink--type structures including a kink with oscillatory structure.

\begin{figure}[!h]
\center
\includegraphics[width=0.5\textwidth]{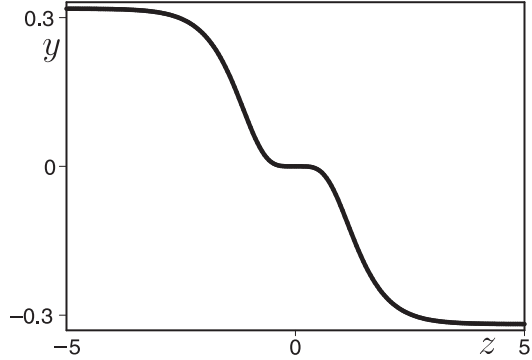}
\caption{Solution \eqref{eq:II_op_1} of Eq. \eqref{eq:II_op} at $a=10$ and $z_{0}=0$.  }
\label{f1a}
\end{figure}

Now we demonstrate an example of Corollary's 1 application. We assume that $\lambda=0$, $\kappa=3/2$ and $f(y)=4/5ay^{-1/5}$, where $a\neq0$ is an arbitrary parameter. Then from \eqref{eq:L1} and \eqref{eq:L1_11} we get
\begin{equation}
y_{zz}+\frac{4a}{5}y^{-1/5}y_{z}-\frac{4a^{2}}{125}y^{3/5}\left(\frac{3}{2}ay^{4/5}-6\right)=0.
\label{eq:II_op}
\end{equation}
Using Corollary 1 and solving corresponding first order differential equation we find a one--parameter family of solutions of  equation \eqref{eq:II_op}
\begin{equation}
y=-\frac{4\sqrt{2}}{a^{5/4}}\tanh^{5}\left\{\frac{\sqrt{2}a^{5/4}}{25}(z-z_{0})\right\}.
\label{eq:II_op_1}
\end{equation}
Note that here and below we denote by $z_{0}$ an arbitrary constant. One can see that solution \eqref{eq:II_op_1} describes a kink--type structure and its plot is demonstrated in Fig.\ref{f1a}.

\subsection{Example of Theorem 2 application.}

Now we give an example of application of Theorem 2. We suppose that $\mu=a^{2}$, $\nu=1$ and $f(y)=4y^{3}+4ay$, where $a$ is an arbitrary parameter. In this case with the help of \eqref{eq:L1_V_3} we find corresponding Li\'{e}nard equation
\begin{equation}
y_{zz}+(4y^{3}+4ay)y_{z}-4\alpha y(y^{2}+a+\alpha)^{2}+16\beta y=0
\label{eq:LE_e2_1}
\end{equation}
The general solution of equation \eqref{eq:LE_e2_1} can be obtained with the help of formulas \eqref{eq:L1_V_1} and has the form
\begin{equation}
y=\pm \sqrt{2w-\alpha-a}, \quad z=\int\frac{1}{4y} d\zeta,
\label{eq:LE_e2_3}
\end{equation}
where $w$ is given by \eqref{eq:L1_V_gs}. Let us remark that we use the upper sign in formulas \eqref{eq:L1_V_3}, \eqref{eq:L1_V_1}.

\begin{figure}[!h]
\center
\includegraphics[width=0.75\textwidth]{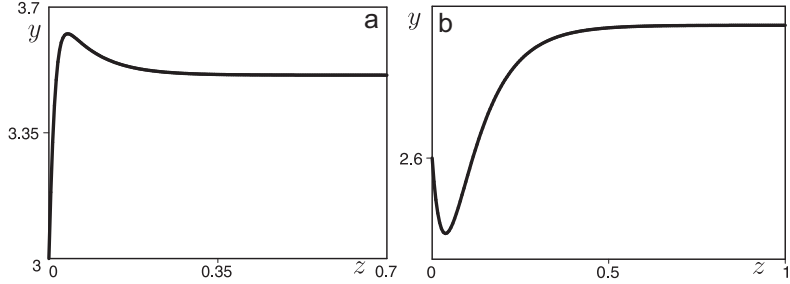}
\caption{Solution \eqref{eq:LE_e2_3} of equation \eqref{eq:LE_e2_1} at: a) $\alpha=-1$, $\beta=-10$ and $a=-5$ ; b) $\alpha=-1$, $\beta=-2$ and $a=-3$.  }
\label{f2}
\end{figure}

We demonstrate plots of solution \eqref{eq:LE_e2_3} corresponding to the plus sing at various values of the parameters in Fig. \ref{f2}. One can see that solution \eqref{eq:LE_e2_3} describes various kink--type structures.

Let us consider an example of Corollary's 2 application. Suppose that $f=e^{-y}(2\alpha-2e^{-y})$, $\nu=1$ and $\mu=\alpha^{2}$. Then we find a Li\'{e}nard equation satisfying
correlation \eqref{eq:L1_V_3}:
\begin{equation}
y_{zz}-2e^{-y}(e^{-y}-\alpha)y_{z}+2e^{-y}(\alpha e^{-2y}-4\beta)=0.
\label{eq:LE_e2_5}
\end{equation}
Substituting values of $f$, $\mu$ and $\nu$ into \eqref{eq:L1_V_5} and solving corresponding differential equation we obtain
\begin{equation}
y=\frac{1}{2\alpha}\left[8\beta(z-z_{0})+\alpha\ln\left\{\frac{1}{4\beta}(\alpha e^{-\frac{8\beta}{\alpha}(z-z_{0})}-1)\right\}\right].
\label{eq:LE_e2_5_1}
\end{equation}
Thus, we find  a one--parameter family of analytical solutions of equation \eqref{eq:LE_e2_5}.

\subsection{Example of Theorem 3 application.}

Now we study applications of Theorem 3. We suppose that $f(y)=(2y+a)/(y^{2}+ay+b)$ and $\eta=1$. Then, using formula \eqref{eq:L1_XXVII_c} we find the following Li\'{e}nard equation
\begin{equation}
y_{zz}+\frac{2y+a}{y^{2}+ay+b}y_{z}-\frac{2y+a}{y^{2}+ay+b}\left(\sigma^{2}\gamma(y^{2}+ay+b)^{2}-1\right)=0.
\label{eq:LE_e3_1}
\end{equation}

\begin{figure}[!h]
\center
\includegraphics[width=0.4\textwidth]{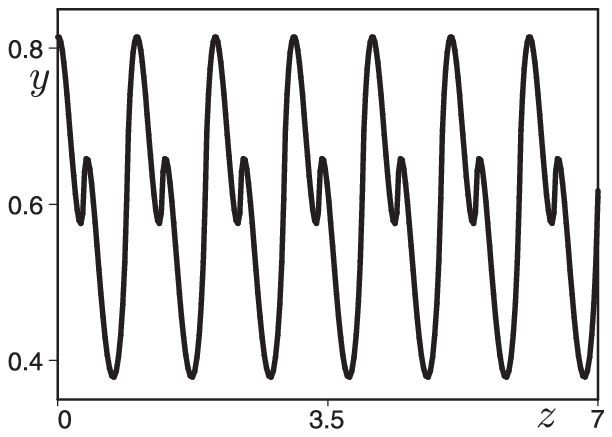}
\caption{ Solution \eqref{eq:LE_e3_3} at $\gamma=-\sigma=-2$, $C_{4}=i$, $\zeta_{0}=1/2$, $a=-b=1$. }
\label{f3}
\end{figure}

The general solution of equation \eqref{eq:LE_e3_1} can be obtained with the help of formulas \eqref{eq:L1_XXVII_gs} and \eqref{eq:L1_XXVII_f}:
\begin{equation}
y=\frac{1}{2}\left(-a\pm \sqrt{\frac{4w}{\sigma}+a^{2}-4b}\right), \quad z=-\int\frac{1}{\sigma(2y+1)}d\zeta,
\label{eq:LE_e3_3}
\end{equation}
where $w$ is given by \eqref{eq:L1_XXVII_gs}. Let us remark that solution \eqref{eq:L1_XXVII_gs}, and therefore \eqref{eq:LE_e3_3} can have a real period in the case of $\gamma<0$. We demonstrate solution \eqref{eq:LE_e3_3} corresponding to the plus sing in \eqref{eq:LE_e3_3} in Fig.\ref{f3}.

Corollary 3 allows us to construct one--parameter families of analytical solutions of equation \eqref{eq:L1} if condition \eqref{eq:L1_XXVII_c} holds. Suppose that $f=\tan y$ and $\eta=-1$, then from \eqref{eq:L1} and \eqref{eq:L1_XXVII_c} we find corresponding Lienard equation
\begin{equation}
y_{zz}+(\tan y) y_{z}+\tan y(\sigma^{2}\gamma \cos^{2}y-1)=0.
\label{eq:LE_e3_5}
\end{equation}
Substituting $\eta=-1$ and $C_{2}=0$ into  \eqref{eq:L1_XXVII_a} and solving corresponding differential equation we get
\begin{equation}
\begin{gathered}
y=2\arctan\left\{\frac{\sqrt{\gamma\sigma^{2}-1}}{\sqrt{\gamma}\sigma-1}\tanh\left[\frac{\sqrt{\gamma\sigma^{2}-1}}{2}(z-z_{0})\right]\right\}, \quad \sigma^{2}\gamma\neq1,
\label{eq:LE_e3_7}
\end{gathered}
\end{equation}
Notice that when $\sigma=\pm 1/\sqrt{\gamma}$ solution \eqref{eq:LE_e3_7} has either the form $y=2\arctan\{z-z_{0}\}$ or  $y=-2\arctan\{1/(z-z_{0})\}$ correspondingly.

\begin{figure}[!h]
\center
\includegraphics[width=0.75\textwidth]{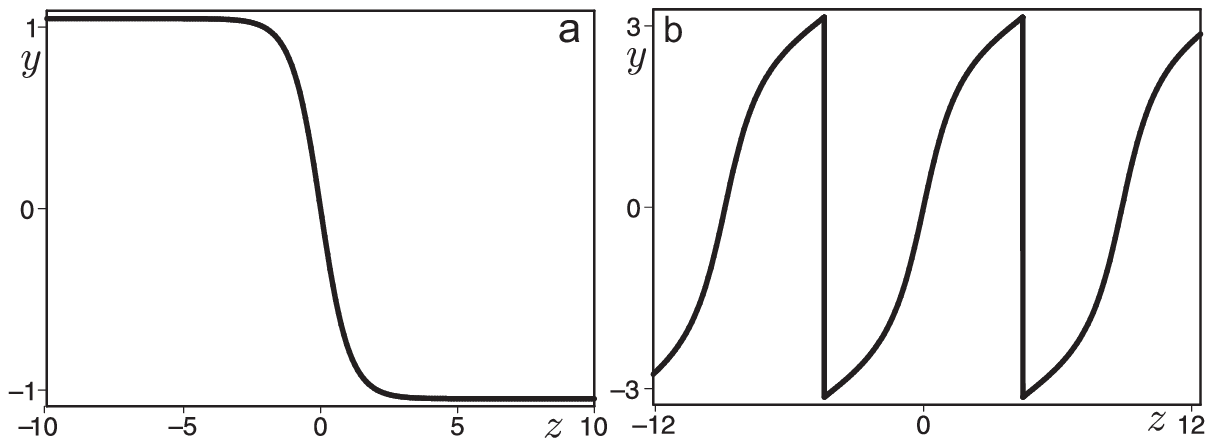}
\caption{ Solution \eqref{eq:LE_e3_7} of equation \eqref{eq:LE_e3_5} at a) $\gamma=1$, $\sigma=-2$ and $z_{0}=0$; b) $\sigma=1$, $\gamma=1/2$ and $z_{0}=0$. }
\label{f3a}
\end{figure}

Thus, we find one--parametric solutions family of \eqref{eq:LE_e3_5}.  When $\gamma\sigma^{2}>1$ solution \eqref{eq:LE_e3_7} is a smooth and monotonic function, while in the case of $\gamma\sigma^{2}<1$ is is a periodic discontinuous function.  We demonstrate plots of solution \eqref{eq:LE_e3_7} for the both cases $\gamma\sigma^{2}>1$ and $\gamma\sigma^{2}<1$  in Fig.\ref{f3a}.

In this section we have constructed three new examples of integrable Li\'{e}nard equations with the help of Theorems 1--3. We have obtained closed--form expressions for the general solutions of these equations. We have also demonstrated that we can find explicit expressions for the one--parametric families of analytical solutions of the Li\'{e}nard equation by means of Corollaries 1--3. We believe that all solutions obtained in this section are new.

\section{Conclusion}

In this work have studied connections between the Li\'{e}nard equation and equation of the Painlev\'{e}--Gambier type. We have considered all subcases of equation \eqref{eq:L1} that belong to the Painlev\'{e}--Gambier classification. We have demonstrated that some of these equation can be linearized via the generalized Sundman transformations, and, therefore do not lead to new integrability conditions for the Li\'{e}nard equation. On the other hand, the rest Painlev\'{e} integrable subcases of equation \eqref{eq:L1} give us new criteria for integrability of the Li\'{e}nard equation. The case of equation VII was considered in \cite{Kudryashov2016}, while connection between equation \eqref{eq:L1} and equations II, V and XVII from the Painlev\'{e}--Gambier classification have been considered in the present work. As a result, we have found three new criteria for the integrability of the Li\'{e}nard equation. We have demonstrated applications of our approach by constructing general analytical solutions of three new integrable Li\'{e}nard equations. We have also proposed an approach for finding one--parameter families of closed form analytical solutions of the Li\'{e}nard equation. We have shown that in some cases we can effectively find one--parametric families of analytical solutions of the Li\'{e}nard equation.

\section{Acknowledgments}
This research was partially supported by grant for the state support of young Russian scientists 6624.2016.1 and by the grant for the state support of scientific schools 6748.2016.1.

\end{document}